# Problems in Systematic Application of Software Metrics and Possible Solution

Gordana Rakić, Zoran Budimac

*Department of Mathematics and Informatics,
Faculty of Sciences,
University of Novi Sad,
Trg Dositeja Obradovića 4,
21000 Novi Sad, Serbia*

`goca@dmi.uns.ac.rs, zjb@dmi.uns.ac.rs`

*Abstract* - Systematic application of software metric techniques can lead to significant improvements of the quality of a final software product. However, there is still the evident lack of wider utilization of software metrics techniques and tools due to many reasons. In this paper we investigate some limitations of contemporary software metrics tools and then propose construction of a new tool that would solve some of the problems. We describe the promising prototype, its internal structure, and then focus on its independency of the input language.

## I. Introduction

Software metric can be defined as measure that reflects some property of a software product or its specification. Software metric value can also be related to only one unit of a software product. There are numerous categorizations of software metrics but considering the measurement, target metrics could be divided in three main categories: product metrics, process metrics and project metrics [9]. In this paper we shall deal with the product metrics and especially code metrics as its sub-category. Although metric values could be calculated manually, nowadays software metric tools are being used for calculation of metric values and for their further processing and analysis. Software metrics and software metric tools are wide research areas and improvements in these fields may bring higher success of software projects in general. However, the state of the art in the field shows that there is no wider acceptance of techniques and therefore still no significant improvements. A new software metrics tool with advanced features would play important role in these improvements.

Motivations behind designing a new tool lay in numerous reports on weaknesses of existing tools both from practice and from academic world (e.g. [12], [13], [15], [17]) of which we enumerate some:
- Software metrics tools are generally not independent on programming language and/or underlying platform. Therefore different tools are often used for different projects, for different components of one project, or even within a single software component.
- Different tools sometimes provide inconsistent results. Therefore, usage of different tools for different software components is not only hard but also often unusable.
- Tools usually compute only a selection of possible metrics. They also rarely combine them to gain higher measure quality and also rarely store the code/metrics values to track changes over time.
- Tools rarely display values in 'user-friendly' way to a non-specialist (e.g., graphically) and rarely interpret the meaning of computed numerical results and their correlations. They almost never suggest what typical actions should be taken in order to improve the quality of the code.
- Tools are typically insensitive to the existence of additional, useless and duplicate code that can be present for tracking, testing and debugging purposes.
- Tools usually do not deal with attempts to 'cheat' the metrics algorithm. Cheating is possible if the internal characteristics of the employed techniques/tools are known - in such cases programmers can spend more time adjusting the code to the tool, rather than reaching the real program quality.
- Sometimes it is not clear which specific software metric has to be applied to accomplish the specific goal. Frequently, the reason for this confusion lies in the gap between the real quality parameters for software projects and strictly defined mathematical functions (that lies in the hearth of software metrics).

TABLE I
THE OVERVIEW OF THE RESULTS OF SOFTWARE METRIC TOOLS REVIEW.

| Tool | Producer [see ref] | Platform independ. | Language independ. | Supported metrics | | | | | Code hist. | Metrics storing. | Graphical represent. | Interpretation /Improvement |
|------|---------|---|---|---|---|---|---|---|---|---|---|---|
| | | | | CC | H | LOC | OO | others | | | | |
| SLOC | D. Wheeler [23] | - | + | - | - | + | - | - | - | + | - | - |
| Code Counter Pro | Geronesoft [7] | - | + | - | - | + | - | - | - | + | - | - |
| Source Monitor | Campwood Software [20] | - | - | + | - | + | + | - | - | + | + | - |
| Understand | ScientificToolworks [22] | + | - | + | + | + | - | - | - | + | - | - |
| RSM | MSquaredTechnologies [18] | + | - | + | + | + | - | - | + | + | - | - |
| Krakatau | Power Software [10], [11] | - | +* | + | + | + | + | - | - | + | + | - |

Developing a software metrics tool that will solve at least some of the enumerated disadvantages would increase the level of application of software metrics in practice and, in so doing, improve the development process and final product quality.

In this paper we shall describe the construction of a new tool that addresses some of the above-mentioned problems. After that we shall focus on the features of the tool that enable its independency on the input language and on metrics algorithms. We consider these features as the starting as well as the most important ones..

The rest of the paper is organized as follows. To justify the need for a new tool, a review of software metric tools and related ideas is given in section 2. In Section 3 the three basic steps in creating the tool are explained. Description of the developed prototype of the new tool follows in section 4. Internal representation of the source code is described in section 5. Small and representative test example is presented in section 6. Conclusions and further work are given in section 7.

II. RELATED WORK

In the first part of this section we analyze tools with respect to two groups of criteria and show that the aim to create a new one is justified. In the second part of the section we are focusing on related work on language independency.

A. *Two groups of general criteria*

The first group of criteria is related to the possibility of wide usage of the tool, independently of nature and structure of the software product that is to be measured. Those are: platform independency, input language independency, and supported metrics.

The second group of criteria is related to keeping and processing produced results and intermediate results. These criteria are: history of code, metric results storing facility, graphical representation of calculated values, and interpretation of calculated values including suggestions for the improvements based on calculated values.

The analysis included 20 tools, out of which six representative examples are displayed in Table I. Other tools (of the 20) belong to the same classes as the displayed ones. Symbol "+" indicates that listed tool possess corresponding characteristic, while "-" indicates that this criterion is not satisfied. Mark "*" next to the symbol "+" means that tool only partially satisfies specified criterion.

The table includes support for the following metrics:
• Cyclomatic Complexity (CC) that reflects structure complexity based on control-flow structures in the program.
• Halstead Metrics (H) that reflects complexity of the program based on number of operators and operands.
• Lines of Code (LOC) that represents length of the source code expressed in number of the lines of the source code. It is common to make difference between number of the lines of comment (CLOC), source code (SLOC), etc. In this analysis, if some tool calculates any of LOC metric, than corresponding cell contains "+" symbol.
• Object Oriented Metrics (OO) – big family of metrics related to object-orientation of the program. If some tool supports any of OO metrics then corresponding cell contains "+" symbol.
• Others – if any metric that does not belong to any of listed categories is supported.

The most important conclusions of the analysis follow:
• Analyzed tools could be divided in two categories. The first category includes tools that calculate only simple metrics as are metrics from LOC family, but for wide set of programming languages. The second category of tools is characterized with wide range of metrics, but limited to a small set of programming languages. There are attempts to bridge the gap between these categories, but without final success. This is the big limitation not only for reasons noted in the introductory section, but also because there are many legacy software systems written in 'ancient' languages to which modern metrics tools cannot be applied uniformly.
- Even if tools support some object-oriented metrics, the amount of supported OO metrics is fairly small. Especially comparing to the broad application of the object-oriented approach in current software development.
- Many techniques/tools compute numerical results with no real interpretation of their meaning. The only interpretation of numerical results which can be found is graphical. These results possess little or no value to practitioners who need suggestion or advice how to improve their project based on metrics' results.

### B. Input language independency

In this section we are focusing on various software tools that strive to achieve independency on the input (programming) language.

ATHENA tool for assessing the quality of software [3] was based on the parsers that generate abstract syntax trees as a representation of the source code. Generated trees were structures that the metric algorithms were applied to. Final goal of these calculations was to generate a report on product quality. This tool was only executable under UNIX operating system. More details about this tool as well as information about its further development are not available anymore.

Development of the CodeSquale metrics tool was based on similar idea. Early results were published on the project website [4, 5]. They developed a system based on representation of a source code through abstract syntax trees, and one object oriented metric for Java source code was implemented. Also, the idea for the further implementation of the other metrics and opportunities for extension of the implementation to the other programming languages were described. Stated final goal was programming language independency. Unfortunately, later results were not published.

Static analysis usually includes some metrics calculation and further analysis of the obtained values. Static analysis of student programs written in Java is the subject of [21]. It is based on usage of abstract syntax tree (AST) to represent the code, but in this case in the XML format.

AST representation of the source code led to language independency in some related areas of source code analysis. Some of the examples follow, but it should be noticed that the nature of referenced problems makes the language independency goal more reachable.

The tool described in [6] uses the abstract syntax tree for representation of the source code in duplicated code analysis. The tree has some additional features designed for easier implementation of the algorithm for comparison.

A similar approach to the code clone analysis by AST representation of the source code was described in [2], but the more complex algorithm for comparison was implemented.

To detect similar classes in Java code, analogous approach was used in [19].

Furthermore, ASTs were used for monitoring of changes [14]. Specified tool was implemented for the analysis of code written in programming language C. Its significance is in mentioning useful ideas for change analysis based on AST.

### III. TOWARDS THE TOOL

Having the enumerated flaws in mind (sections 1 and 2), our new software metrics tool is being constructed to remedy some of the flaws by achieving the following goals:
- Platform independency
- Programming language independency
- Applicability to broad spectre of metrics
- Storing the source code history
- Storing the calculated metrics values
- Interpretation of metric results to the end user
- Improvement recommendations to the user

The basic idea was to split complete development of described tool into three steps with explicit goals for each step (Figure 1):

- **Step 1** - to create an appropriate intermediate structure for the representation of a source code in various input languages and to which software metrics algorithms can be applied.

- **Step 2** - to convert input languages to the structure; to apply software metrics algorithms to the given structure; to produce appropriate numerical values as a result.

- **Step 3** - to apply advanced algorithms to the values of metrics calculated in step 2, in order to produce more usable information to the end user.

Each step could be divided in corresponding number of sub-steps in order to meet some specific goal, e.g. to deal with duplicate or dead code in step 2 or to prevent developers' cheating in step 3.

### A. Step 1 – Intermediate Structure

Since our tool should be language independent it is required to create a special intermediate structure for program representation. As our review in section 2.2 showed, AST could be a good basis for reaching that objective – language independency. AST can represent all elements of the source code.

Software metrics are more sensitive to input programming language syntax than mentioned related fields. Therefore, our approach is to use Concrete Syntax Trees (CSTs) instead of ASTs. CST represents concrete source code elements attached to corresponding language structures.

Although CSTs contain all information about language structures and elements of the source code, they are "too-unaware" of semantic issues of the syntax. In other words, semantically equivalent syntax constructions of two different languages are represented differently (e.g. 'elsif' and 'else if' - see also section 5). Therefore, application of metrics algorithms to this structure requires modification of the CST to avoid separate implementation of metric algorithms for each language. We use slightly enriched form of CST, so-called enriched Concrete Syntax Tree (eCST). eCSTs are stored in suitable XML structure (section 5). Enriched Concrete Syntax Tree (eCST) represents concrete source code elements attached to corresponding language elements, but also contains additional information stored as universal nodes as markers for language elements figuring in metric algorithms.

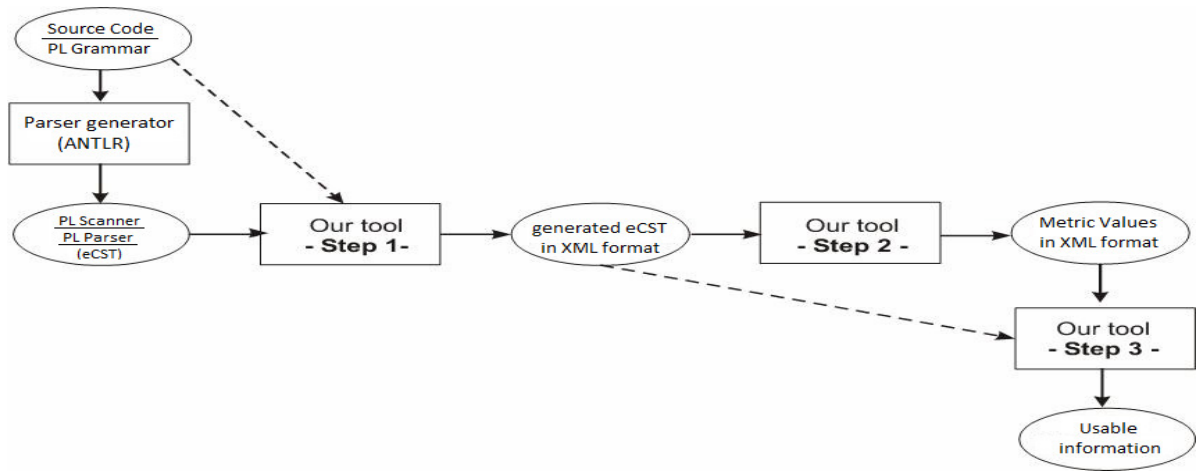

Fig 1. Tool development roadmap

To produce eCST we start from some existing parser generator which as input expects a language grammar and as output returns the language scanner and parser. Besides that, parser generators are usually able to embed into generated translators a mechanism for generating ASTs and CSTs as intermediate structures [8].

We are using ANTLR parser generator [1], [16]. ANTLR accepts language grammar as its input and produces language scanner, parser, AST and CST. CSTs can be extended to eCSTs with additional (imaginary) nodes and enriched with additional information. This enrichment is possible by simple changes inserted in the language grammars that affect only content of the generated tree. Definition of structure of syntax trees is always the same and it is determined by used parser generator (Section 5.1).

Let us note that the usage of other parser generators is also possible as long as they can generate CSTs in a similar form and have possibility of adding new nodes to them.

### B. Step 2 - Calculating Metrics Values

eCST of the source code is the starting point for implementation of as many metric algorithms as possible and to produce rich enough set of numerical characteristics of the source code, needed by the third step. The set of metrics which is to be calculated consists of code metrics, but also all other categories of metrics which could be calculated on a source code. Special attention at this point is given to the application of object oriented metrics.

Also, the decision about the technique that has to be chosen for storing and keeping calculated data was made in this step. Calculated data have to be well organized and prepared for further manipulation necessary in the third step.

### C. Step 3 - Usable Information

After application of all metrics and collecting necessary values, calculated data should be input parameters to advanced algorithms and heuristics for delivering useful information to the end user. Important part of the algorithm is filtering of metric values by applicability to certain programming paradigms. Provided information may be in the form of practical advice for improving the source code, design, project, etc.

## IV. THE PROTOTYPE

The prototype of the new software metrics tool has been implemented in Java. It dynamically recognizes input programming language based on extension in the file name. After that it reads necessary information (on language scanner, language parser, additional information) from a simple XML file that keeps information on all supported languages. After scanner and parser classes have been detected source code is parsed, eCST is generated, and stored to XML file (section 5).

In the next stage eCST in XML format is parsed. During this, metric values are calculated and results are stored to new XML file. XML structure for storing of metric values contains brief information about corresponding elements of the source code. Described architecture of the prototype is presented in figure 2 and flow diagram is shown in figure 3

Our prototype currently supports two characteristic programming languages: procedural Modula-2 and object-oriented Java, and two characteristic metrics: 'physical' LOC and structural CC.

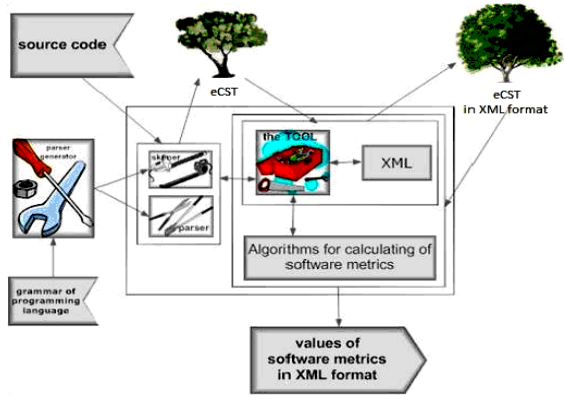

Figure 2. Tool architecture

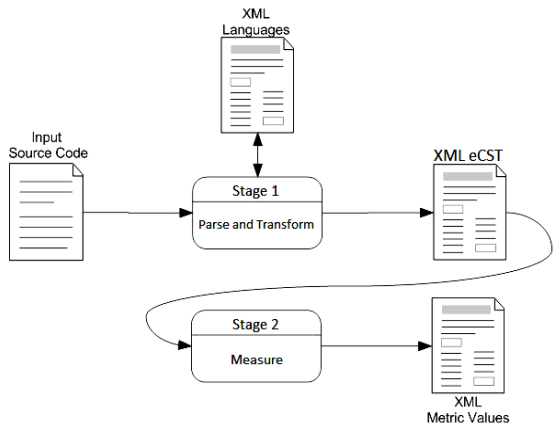

Fig 3. Flow diagram

## I.  eCST - INTERMEDIATE STRUCTURE

Our eCST is meant to be 'universal' tree-structure suitable for representation of the source code and implementation of software metrics algorithms. It is based on:
- comparative analysis of applications of one metric to different programming languages;
- comparative analysis of applications of different metrics to one programming language.

Determined structure is suitable for unique representation of a source code written in different programming languages but, also for an application of different metrics algorithms.

For example, one of the CC calculation algorithms is based on counting certain predicates indicating loops, branches, logical operations, etc. These predicates are usually different in different programming languages. This is the reason for adding unique imaginary node before each branch, each loop, etc. which will initiate recognition and counting of the important elements of language syntax independently of programming language. eCST generated in that way enables trouble-free implementation of metrics calculation algorithms (section 5.2).

### A.  Storing eCST

Generated eCST consists of nodes and branches. Some of the nodes are "imaginary" and provide useful additional information about structure of the source code and input programming language elements. Each node consists of general data about characteristics and position of the source code element and possible sub-nodes. This is the basic structure of syntax trees. Parsers generated by different parser generators are usually producing trees in that or in slightly modified form.

XML schema for keeping generated eCST is presented in figure 4.

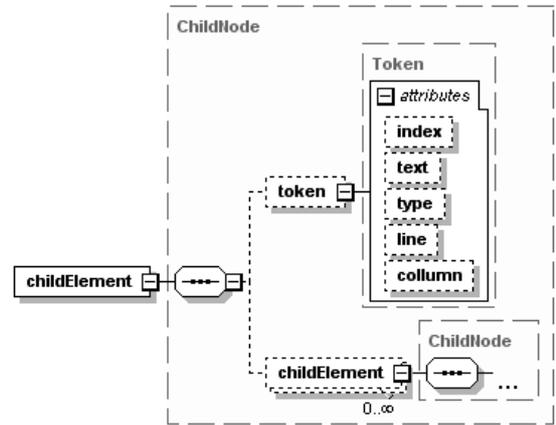

Figure 4. eCST XML schema

### B.  Example – imaginary nodes

The following example shows how the simple "if" statement is represented by CST and eCST and stored to the given XML structure. The statement

```
IF a > b THEN res = 1;
ELSE IF a = b THEN res = 0;
ELSE rez = -1;
```

is represented as CST and eCST as displayed in figures 5 and 6, while the corresponding grammar rule is given in Source Code 1.

SOURCE CODE I

GRAMMAR RULE EXAMPLE - IF STATEMENT

```
IF parenthesizedExpression ifStat = statement
    (ELSE elseStat = statement
    ->  ^(BRANCH_STATEMENT IF   ^(BRANCH ^(IF parenthesizedExpression $ifStat))
     ^(BRANCH ELSE $elseStat))|
    ->  ^(BRANCH_STATEMENT IF ^(BRANCH ^(IF parenthesizedExpression $ifStat)))
    )
```

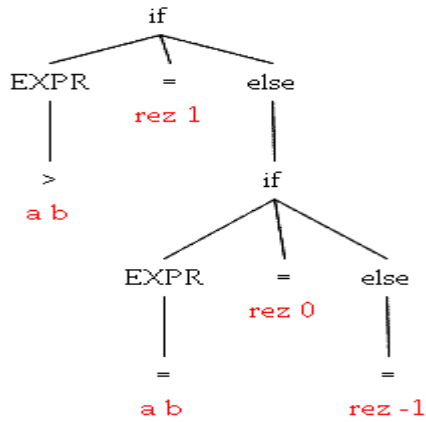

Fig 5. CST - IF statement

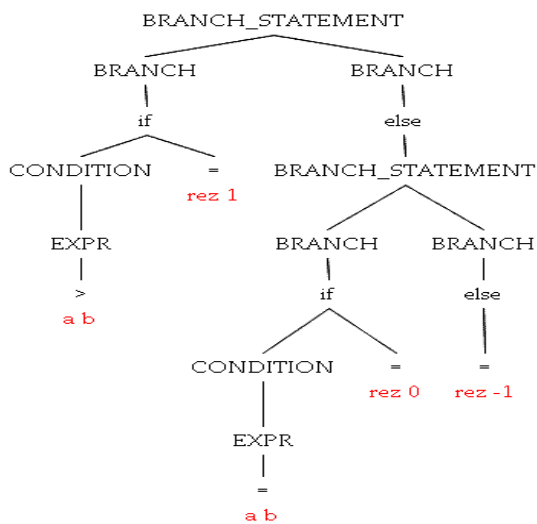

Fig 6. eCST - IF statment

Note that eCST contains some imaginary nodes (BRANCH_STATEMENT, BRANCH, CONDITION, etc). These nodes are added to the eCST by adding corresponding elements in the grammar rule (bolded) to achieve language independency. In this particular case, adding these three nodes lead to the language independency in calculating of CC metric. "BRANCH_STATEMENT" represents the beginning of the whole block that contains "if" statement. It may consist of one or more sub-trees whose root is node named "BRANCH" and represents start of the each branch in the whole branching block. Furthermore, each sub-tree that contains single branch may contain sub-tree representing condition. Root of this sub-tree is "CONDITION" node.

## V. TEST EXAMPLE

The work of the prototype is illustrated on the Quick Sort algorithm that was implemented in Java and Modula-2 (Source Code 2). Results gained from the prototype for the test case are shown in Table 2

As could be expected, results for LOC metric are different for these two implementations, but values for CC are identical. It should be noted that REPEAT loop in Modula-2 and DO-WHILE loop in Java programming language are recognized as loops independently of different syntax. Also, important point is that WHILE keyword in Java language increase value of CC when it is part WHILE loop. This is not the case when WHILE takes part in DO-WHILE loop. This would not be possible without imaginary nodes.

SOURCE CODE II

QUICKSORT algorithm implemented in MODULA-2 and JAVA programming language

| QuickSort.mod<br>**PROCEDURE Sort**(VAR array : ArrayType;<br>          Left, Right : Index**);…**<br>**BEGIN**<br>**…**<br>   **REPEAT**<br>      **WHILE** (array[i] < Middle ) **DO**<br>         INC(i);<br>      **END**;<br>      **WHILE** (array[j] > Middle) **DO**<br>         DEC(j);<br>      **END**;<br>      **IF** (i <= j) **THEN**<br>         …<br>      **END**;<br>   **UNTIL** (i > j);<br>   **IF** (Left < j) **THEN**<br>      Sort(array, Left, j);<br>   END;<br>   **IF** (i < Right) **THEN**<br>         Sort(array, i, Right);<br>   **END**<br>**END Sort;** | QuickSort.java<br>**public static void sort(**<br>          int[] array,int left,int right)<br>{<br>…<br>   do{<br>      while (array[i] < middle)<br>         i++;<br><br>      while (array[j] > middle)<br>         j--;<br><br>      if (i <= j) {<br>         …<br>      }<br>   }while (i <= j);<br>   if (left < j)<br>      sort(array, left, j);<br><br>   if (i < right)<br>      sort(array, i, right);<br><br>} |

ICIT 2011  The 5th International Conference on Information TechnologyTABLE II

SOFTWARE METRICS RESULTS

| MODULA-2 | | | | JAVA | | | |
|---|---|---|---|---|---|---|---|
| PL element | Anotation in eCST | CC | LOC | PL element | Anotation in eCST | CC | LOC |
| *Sort* | *FUNCTION_DECL* | 7 | 32 | *Sort* | *FUNCTION_DECL* | 7 | 27 |
| REPEAT | LOOP_STATEMENT | 4 | 15 | DO-WHILE | LOOP_STATEMENT | 4 | 15 |
| WHILE | LOOP_STATEMENT | 1 | 3 | WHILE | LOOP_STATEMENT | 1 | 2 |
| WHILE | LOOP_STATEMENT | 1 | 3 | WHILE | LOOP_STATEMENT | 1 | 2 |
| *BRANCHING* | BRANCH_STATEMENT | 1 | 7 | *BRANCHING* | BRANCH_STATEMENT | 1 | 7 |
| IF | BRANCH | 1 | 6 | IF | BRANCH | 1 | 7 |
| *BRANCHING* | BRANCH_STATEMENT | 1 | 3 | *BRANCHING* | BRANCH_STATEMENT | 1 | 2 |
| IF | BRANCH | 1 | 2 | IF | BRANCH | 1 | 2 |
| *BRANCHING* | BRANCH_STATEMENT | 1 | 3 | *BRANCHING* | BRANCH_STATEMENT | 1 | 2 |
| IF | BRANCH | 1 | 2 | IF | BRANCH | 1 | 2 |

## VI. CONCLUSION AND FURTHER WORK

In this paper the most important weaknesses in the area of software metrics and tools have been examined and presented together with possible solutions. The basic idea for development of a new software metrics tool has been proposed and the first prototype of the new software metrics tool has been described.

The prototype was developed in Java with the intention to be platform independent. Prototype is built around our own extended concrete syntax tree (eCST) that serves as an intermediate structure suitable for representation of various programming languages and for being processed by various software metrics algorithms. Thus we would achieve a 'universal' software metrics tool – the goal that has not been reached so far, but is needed by many practitioners and software assurance officers.

We are currently at step 2 of the general plan (see section 3) where the implementation for other languages and metrics is to be developed. Special attention is paid to 'ancient' programming languages (e.g. COBOL and FORTRAN) for which we are currently developing eCSTs.

Furthermore relations between compilation units are not yet considered. Prototype currently analyses single compilation unit stored in a single file.

## ACKNOWLEDGEMENTS

Bilateral project between Serbia and Slovenia (project no. 27, 2010-2011) enabled the exchange of visits and ideas with colleagues of Faculty of Electronics, Computing and Informatics (Maribor, Slovenia).## REFERENCES

[1] *ANTLR*, 2010, http://www.antlr.org
[2] I.D.Baxter, A. Yahin, L. Moura, M. Sant'Anna, L. Bier, *Clone Detection Using Abstract Syntax Trees*, Proc. of International Conference on Software Maintenance, 1998. ISBN: 0-8186-8779-7, pp. 368-377
[3] D.N. Christodoulakis, C. Tsalidis, C.J.M. van Gogh, V.W. Stinesen, *Towards an automated tool for Software Certification*, International Workshop on Tools for Artificial Intelligence, 1989. Architectures, Languages and Algorithms, IEEE, ISBN: 0-8186-1984-8, pp. 670-676
[4] *CodeSquale* , 2010, http://code.google.com/p/codesquale/
[5] *CodeSquale*, 2010, http://codesquale.googlepages.com/
[6] S. Ducasse, M. Rieger, S. Demeyer, *A Language Independent Approach for Detecting Duplicated Code*, Proc. IEEE International Conference on Software Maintenance, 1999. (ICSM '99) , ISBN: 0-7695-0016-1, pp 109-118
[7] *Code Counter Pro*, 2010, http://www.geronesoft.com/
[8] Grune D., Bal H.E., Jacobs C.J.H., Langendoen K.G., *Modern compiler design*, – John Wiley & Sons, England, 2000, ISBN 0-471-97697-0, 753p
[9] S. Kan, *Metrics and Models in Software Quality Engineering – Second Edition*, Addison-Wesley, Boston, 2003, ISBN 0-201-72915-6
[10] *Krakatau Suite Management Overview,* , 2010, http://www.powersoftware.com/
[11] *Krakatau Essential PM (KEPM)- User guide 1.11.0.0*, , 2010, http://www.powersoftware.com/
[12] M.Lanza, R.Marinescu, *Object-Oriented Metrics in Practice – Using Metrics to Characterize, Evaluate, and Improve the Design of Object-Oriented Systems*, Springer – Verlag, Berlin, Heidelberg, New York, Germany, 2006, ISBN 978-3-540-24429-5
[13] R. Lincke, J. Lundberg, W. Löwe, *Comparing software metrics tools,* Proc. of the 2008 international symposium on Software testing and analysis ISSTA '08, pp. 131-142
[14] Neamtiu, J. S. Foster, M. Hicks, *Understanding source code evolution using abstract syntax tree matching*, Proceedings of the International Conference on Software Engineering  2005, international workshop on Mining software repositories, ISBN:1-59593-123-6, pp 1–5
[15] J. Novak, G. Rakić, *Comparison of Software Metrics Tools for :NET*, Proc. 13th International Multiconference Information Society - IS 2010, vol. A, pp. 231-234
[16] Parr T., *The Definitive ANTLR Reference - Building Domain-Specific Languages*, The Pragmatic Bookshelf, USA, 2007, ISBN: 0-9787392-5-6
[17] Rakić G., Budimac Z., Bothe K*., Towards a 'Universal' Software Metrics Tool-Motivation, Process and a Prototype*,6